\documentclass[12pt,english,floatfix,superscriptaddress,aps,prd,preprint,showkeys]{revtex4}

\usepackage[utf8]{inputenc}
\RequirePackage{flushend}
\usepackage{array}
\usepackage{lipsum}
\usepackage{graphicx}
\usepackage{amsmath,amsthm,amsfonts,amssymb}
\usepackage{graphicx}
\usepackage[english]{babel}
\usepackage{color}
\usepackage[dvips]{epsfig}
\usepackage[dvips]{graphicx}
\usepackage{float}
\usepackage{units}
\usepackage{textcomp}
\usepackage{mathrsfs}
\usepackage[makeroom]{cancel}

\usepackage{hyperref}
\hypersetup{colorlinks,breaklinks,
			citecolor=[rgb]{0,0.0,1.0},
            urlcolor=[rgb]{0.0,0.0,1.0},
            linkcolor=[rgb]{0,0.5,0.9}}
\usepackage{slashed}
\usepackage{slashed}

\newcommand{\ie}{\begin{equation}}
\newcommand{\fe}{\end{equation}}
\newcommand{\se}{\begin{eqnarray}}
\newcommand{\ff}{\end{eqnarray}}

\begin{document}

\title{The Casimir effect for the scalar and Elko fields in a Lifshitz-like field theory} 


\author{R. V. Maluf}
\email{r.v.maluf@fisica.ufc.br}
\affiliation{Universidade Federal do Cear\'a (UFC), Departamento de F\'isica,\\ Campus do Pici, Fortaleza,  CE, C.P. 6030, 60455-760 - Brazil.}

\author{D. M. Dantas}
\email{davi@fisica.ufc.br}
\affiliation{Universidade Federal do Cear\'a (UFC), Departamento de F\'isica,\\ Campus do Pici, Fortaleza,  CE, C.P. 6030, 60455-760 - Brazil.}

\author{C. A. S. Almeida}
\email{carlos@fisica.ufc.br}
\affiliation{Universidade Federal do Cear\'a (UFC), Departamento de F\'isica,\\ Campus do Pici, Fortaleza,  CE, C.P. 6030, 60455-760 - Brazil.}

\date{\today}


\begin{abstract}
In this work, we obtain the Casimir energy for the real scalar field and the Elko neutral spinor field  in a field theory at a Lifshitz fixed point (LP). We analyze the massless and the massive case for both fields using dimensional regularization. We obtain the Casimir energy in terms of the dimensional parameter and the LP parameter. Particularizing our result, we can recover the usual results without LP parameter in (3+1) dimensions presented in the literature. Moreover, we compute the effects of the LP parameter in the thermal corrections for the massless scalar field.
\end{abstract}



\maketitle

\section{Introduction}

The Casimir Effect is characterized by force between two neutral parallel conducting plates, separated by a very short distance,  in the vacuum \cite{Casimir}.  This effect was conceived by Hendrik Casimir, in 1948. The Casimir Effect arises by the difference in the vacuum expectation value of the energy due to the quantization of the electromagnetic field \cite{Bordag, Bordag2}. Experimentally, the Casimir effect was verified at the micrometer scale by \textup{Al} plates in 1958 \cite{Casimir-exp0}, and layers of \textup{Cu} and \textup{Au} in 1997 \cite{Casimir-exp}. A modern review of experimental methods, issues, precisions and realistic measures is presented in \cite{Casimir-exp2, Casimir-exp3}. Once that several factors can modify the zero energy, the Casimir force depends on many parameters: as the geometry and the boundary \cite{Stokes,  c1, c2}, the type of field studied \cite{Stokes, Pereira, Scalar}, and the presence of extra-dimensions \cite{Ponton}. Moreover, some results of Casimir force were applied to black holes \cite{c1}, cosmic strings \cite{c2}, modified gravity \cite{MD}, Lifshitz-like field theory \cite{Petrov, Petrov2, c3} and others Lorentz Violation scenarios \cite{LV}.  In particular, the scalar field was studied in the context of Casimir force in several works \cite{Petrov, Petrov2, Petrov3, c3, Amb, B1, KimballBook, chineses, Report1986}, which will be detailed along with all the paper.

On the other hand, the so-called Elko field (dual-helicity eigenspinors of the charge conjugation operator) are $1/2$ spin neutral fermion field with mass dimension one in $(3+1)$ dimensions \cite{Ahluwalia1, Ahluwalia2, Ahluwalia3, 2019a,  elko-rot}. Due to this, the interactions of Elko spinor field is restricted to only the gravity field and the Higgs field. Hence these fields are dark matter candidates \cite{elko-dk1,elko-dk2}. Besides, Elko fields have several recent applications to the  cosmological inflation \cite{cosmo-1,cosmo-2,cosmo-3}, Very Special Relativity (VSR) \cite{vsr}, Hawking radiation \cite{hawking} and braneworlds \cite{brane-1, brane-2}. Moreover,  some phenomenological constraints and mechanisms to detect Elko at the LHC have been proposed \cite{elko-lhc1,elko-lhc2}. The influence of Elko in the Casimir Effect on $(3+1)$ spacetime was studied in Ref. \cite{Pereira}, where the Casimir force differs both those found for the scalar field as the Dirac field. The Casimir force is repulsive for the Elko field because the Elko fields are spinors with anti-commutation relation \cite{Pereira}.

Furthermore, from the point of view of modified gravity, the Ho$\breve{{\rm r}}$ava-Lifshitz (HL) theory allows the gravity to be power-counting renormalizable \cite{Hor}. The HL theory introduces a space-time anisotropy by the scaling of coordinates such as $x^{i}\to b x^{i}$ and $t\to b^{\xi}t$, being $b$ a length constant and $\xi$ the critical exponent \cite{Hor}. Moreover, due to the high-derivative been present in the spatial direction only, the quadratic terms improve the ultra-violet (UV) behavior of the particle propagator, without introducing ghostlike degrees of freedom \cite{Q1a}. The {absence} of temporal high-derivative also prevents issues about unitarity breaking, and the spatial high-derivative terms were shown to arise as quantum correction \cite{Q1b}. These high-derivative features  collaborate for the renormalization of theories. The HL proposal induces a Lorentz violation at high energies. However, the usual Lorentz symmetry is preserved at low energies \cite{Hor,Petrov}. The HL theory in the curved space has several applications, for example,  in cosmology \cite{HC1,HC2,HC3}, black holes \cite{HBH,HBH2}, gravitational waves \cite{HBH,HGW1,HGW2}, and black body radiation \cite{HBB}.   Alternatively, when gravity effects are not considered, a field theory at a Lifshitz fixed point (LP) in the flat spacetime has some applications to the Casimir effect \cite{Petrov, Petrov2}. The Casimir force was studied under the influence of the LP field theory for the massless scalar field in Refs. \cite{Petrov, Petrov2}. In this LP field theory, the critical exponent changes the usual result allowing an attractive, repulsive, and even a null Casimir force in $(3+1)$-dimensions.

In this work, we deal with the Casimir force due to a massive real scalar field and the  Elko spinor field in a flat $(3+1)$ LP scenario. We will note that the Casimir energy for the Elko differs from the scalar field one only by a multiplicative factor. In order to regularize the Casimir energy, we apply the dimensional regularization. Hence the Casimir force will be obtained in terms of main parameters: dimension $d$,  mass $m$ and the LP critical exponent $\xi$. Due to the complexity of Casimir energy, we will detail our results in terms of particular choices of the mass and $\xi$,  where some unusual behaviors will be discussed. Moreover, we consider the influence of the LP parameter in the thermal corrections of the Casimir effect.

This paper follows the structure: in Sec. \ref{sec-hl-casimir}, we introduce the massive scalar field and the Elko fermion into a field theory at a Lifshitz fixed point. In Sec. \ref{sec-a}, we present our results: In subsection \ref{sec-massless} we will obtain the results for the Casimir energy for the massless Elko. We will recover both the results for the usual massless scalar field in the LP scenarios \cite{Petrov}, such as the particular massless case of Elko presented in Ref. \cite{Pereira}. In subsection \ref{sec-mass}, we study the massive cases. The result for the massive scalar field and Elko field will be recovered for $\xi=1$. Moreover, the first non-trivial result to LP with $\xi=2$ for the massive scalar field and Elko field will be studied, and also the small-mass regime for all $\xi$. In Sec. \ref{sec-temp}, the role of the  LP parameter in the thermal corrections for the massless scalar field is discussed. The paper finishes in section \ref{sec-conc}, where the main results are summarized.


\section{Casimir energy in a Lifshitz-like field theory}\label{sec-hl-casimir}

In this paper, we study some models in field theories at a Lifshitz fixed point (LP) on flat spacetime, were the critical exponent $\xi$ changes the usual behavior of the Casimir energy. For the massless real scalar field, the result present in \cite{Petrov, Petrov2} points out that the Casimir effect depends on the parameter $\xi$ (allowing the null, attractive or repulsive force, that always decays with the distance). Our main goal is to generalize this result for the massive scalar field (and the Elko field), studying finite-temperature, as well. The influence of the mass parameter will yield unusual results for the Casimir force under LP field theory, as an increasing Casimir force with the distance for larger $\xi$.

Let us start computing the Casimir energy for a free massive real scalar field $\phi$ in LP field theory through the following action in natural units $\hbar=c=1$:
\ie
S=\frac{1}{2}\int dtd^{D-1}x\left(\partial_{0}\phi\partial_{0}\phi-\ell^{2(\xi-1)}\partial_{i_{1}}\partial_{i_{2}}\cdots\partial_{i_{\xi}}\phi\partial_{i_{1}}\partial_{i_{2}}\cdots\partial_{i_{\xi}}\phi-m^{2}\phi^2\right),\label{eq:1}
\fe 
with repeated Latin indices summed from one to $D-1$, and $\xi$ as the above mentioned critical exponent (which will be considered in this work as a positive integer $\xi=1,2,3...$). {The field $\phi(x)$ has a mass dimension $(D-1)/(2\xi)-1/2$ and the usual flat spacetime is recovered when $D=4$ and $\xi=1$}. As required by consistency, the constant $\ell$ has the dimension of length.

The corresponding equation of motion is
\ie
\left(\partial_{0}^{2}+\ell^{2(\xi-1)}(-1)^{\xi}\partial_{i_{1}}\cdots\partial_{i_{\xi}}\partial_{i_{1}}\cdots\partial_{i_{\xi}}+m^{2}\right)\phi(x)=0.
 \fe 

In $(3+1)$-dimension, the operator $\partial_{i_{1}}\cdots\partial_{i_{\xi}}\partial_{i_{1}}\cdots\partial_{i_{\xi}}$ takes the following form:
\ie
\partial_{i_{1}}\cdots\partial_{i_{\xi}}\partial_{i_{1}}\cdots\partial_{i_{\xi}}=\left(\partial_{x}^{2}+\partial_{y}^{2}+\partial_{z}^{2}\right)^{\xi},
\fe which reads,
\ie
\left[\partial_{0}^{2}+\ell^{2(\xi-1)}(-1)^{\xi}\left(\partial_{x}^{2}+\partial_{y}^{2}+\partial_{z}^{2}\right)^{\xi}+m^{2}\right]\phi(x)=0.\label{eq:2}
\fe

For two large parallel plates with area $L^{2}$ and separated by an orthogonal $z$ axis with a small distance $a$ ($a\ll L$), the Dirichlet boundary conditions read
\ie
\phi(x)_{z=0}=\phi(x)_{z=a}=0.\label{eq:3}
\fe

Adopting the standard procedure \cite{Ryder}, the quantum field can be written as
\ie
\hat{\phi}(x)=\sum\limits _{n=1}^{\infty}\sqrt{\frac{2}{a}}\int\frac{d^{2}k}{(2\pi)^{2}}\frac{1}{2k_{0}}\sin\left(\frac{n\pi z}{a}\right)\left[a_{n}({\bf k})e^{-ikx}+a_{n}^{\dagger}({\bf k})e^{ikx}\right],\label{eq:4}
\fe
where $n$ is an integer and we have defined $kx\equiv k_{0}x_{0}-k_{x}x-k_{y}y$, being

\ie
k_{0}\equiv\omega_{n,\xi}({\bf k})= \ell^{\xi-1}\sqrt{\mu^{2\xi}+\left(k_{x}^{2}+k_{y}^{2}+\left(\frac{n\pi}{a}\right)^{2}\right)^{\xi}}, \label{freq}
\fe with $\mu^{\xi}\equiv m\ell^{1-\xi}$. The annihilation and creation operators $a_{n}({\bf k})$ and $a_{n}^{\dagger}({\bf k})$ obey the following commutation relations
\begin{eqnarray}
&&\left[a_{n}({\bf k}),a_{n'}^{\dagger}({\bf k'})\right]=(2\pi)^{2}2\omega_{n,\xi}({\bf k})\delta^{(2)}({\bf k}-{\bf k}')\delta_{nn'},  \nonumber \\ 
&&\left[a_{n}({\bf k}),a_{n'}({\bf k'})\right]=\left[a_{n}^{\dagger}({\bf k}),a_{n'}^{\dagger}({\bf k'})\right] = 0.
\end{eqnarray}

Hence, the Hamiltonian operator $\hat{H}$, resulting from canonical quantization reads\ie
\hat{H}=\sum_{n=1}^{\infty}\int \frac{d^{2}k}{(2\pi)^{2}}\frac{1}{2}\left(a_{n}^{\dagger}({\bf k})a_{n}({\bf k})+L^{2}\omega_{n,\xi}({\bf k})\right). \label{Hamiltonian-S}
\fe 
This expression leads to the vacuum energy for the massive real scalar field {in} the {LP} scenario given by
\ie
E=\langle 0|\hat{H}|0\rangle = \frac{L^{2}}{(2\pi)^{2}}\int 
d^{2}k\sum\limits _{n=1}^{\infty}\frac{1}{2}\omega_{n,\xi}({\bf k}),\label{vacuum}
\fe or more explicitly as
\ie
E=\ \frac{\ell^{\xi-1}L^{2}}{(2\pi)^{2}}\int d^{2}k\sum_{n=1}^{\infty}\frac{1}{2}\sqrt{\mu^{2\xi}+\left(k_{x}^{2}+k_{y}^{2}+\left(\frac{n\pi}{a}\right)^{2}\right)^{\xi}}.
\label{vaccum2}\fe

It is easy to see that the vacuum energy \eqref{vaccum2} is infinite, and thus, some renormalization procedure must be applied to remove the divergences. Before studying the regularization of the Casimir energy \eqref{vaccum2}, let us obtain the Casimir energy for the so-called Elko (the German acronym for  ``eigenspinors of the charge conjugation operator''). {The Elko field is a $1/2$ spin neutral fermion field with mass dimension one in $D=4$.} The Casimir effect for the massive Elko field in the usual flat spacetime was already obtained in Ref. \cite{Pereira}. In this work, we perform a comparative of the massive Elko spinor and the massive scalar field in the LP field theory.

The action for the free massive Elko field $\eta(x)$  in {Lifshitz fixed point theory} is 
\ie
S=\int dtd^{D-1}x\left(\partial_{0}\stackrel{\neg}{\eta}\partial_{0}\eta-\ell^{2(\xi-1)}\partial_{i_{1}}\partial_{i_{2}}\cdots\partial_{i_{\xi}}\stackrel{\neg}{\eta}\partial_{i_{1}}\partial_{i_{2}}\cdots\partial_{i_{\xi}}\eta-m^{2}\stackrel{\neg}{\eta}\eta\right),\label{eq:1e}
\fe 
where the $\eta(x)$ represents the Elko (and its dual $\stackrel{\neg}{\eta}(x)$). 
{Notice that just like the scalar field $\phi(x)$, the field $\eta(x)$ also has a mass dimension $(D-1)/(2\xi)-1/2$}.

Again, the Elko equation of motion in $(3+1)$-dimension is similar to equation \eqref{eq:2}
\ie
\left[\partial_{0}^{2}+\ell^{2(\xi-1)}(-1)^{\xi}\left(\partial_{x}^{2}+\partial_{y}^{2}+\partial_{z}^{2}\right)^{\xi}+m^{2}\right]\eta(x)=0.\label{eq:2e}
\fe

Furthermore, with the Dirichlet boundary conditions
\ie
\eta(x)_{z=0}=\eta(x)_{z=a}=0, \quad \stackrel{\neg}{\eta}(x)_{z=0}=\stackrel{\neg}{\eta}(x)_{z=a}=0,\label{eq:3e}
\fe the Elko quantum field operator stands for
\ie
\eta(x)=\sum\limits _{n=1}^{\infty}\sqrt{\frac{2}{a}}\int\frac{d^{2}k}{(2\pi)^{2}}\frac{\sin\left(n\pi z/a\right)}{\sqrt{2m k_{0}}}\sum_{\beta}\left[a_{\beta,n}({\bf k})\lambda^{S}_{\beta}({\bf k})e^{-ikx}+a_{\beta,n}^{\dagger}({\bf k})\lambda_{\beta}^{A}({\bf k})e^{ikx}\right],\label{eq:4e}
\fe
and 
\ie
\stackrel{\neg}{\eta}(x)=\sum\limits _{n=1}^{\infty}\sqrt{\frac{2}{a}}\int\frac{d^{2}k}{(2\pi)^{2}}\frac{\sin\left(n\pi z/a\right)}{\sqrt{2m k_{0}}}\sum_{\beta}\left[a_{\beta,n}^{\dagger}({\bf k})\stackrel{\neg}{\lambda}^{S}_{\beta}({\bf k})e^{ikx}+a_{\beta,n}({\bf k})\stackrel{\neg}{\lambda}_{\beta}^{A}({\bf k})e^{-ikx}\right],\label{eq:5e}
\fe with the same frequency $\omega_{n,\xi}({\bf k})$ defined in Eq. \eqref{freq} associated.

The spin one-half eigenspinors, $\lambda_{\beta}^{S/A}({\bf k})$, satisfy the eigenvalue equation $C\lambda_{\beta}^{S/A}({\bf k})=\pm \lambda_{\beta}^{S/A}({\bf k})$, being $C$ the charge conjugation operator and  $\lambda^S$ stands for the  self-conjugate spinor (positive), while  $\lambda^A$  stands for its anti self-conjugate (negative). Moreover, the helicity is represented by $\beta=(\{+,-\}, \{-,+\})$ \cite{Pereira, Ahluwalia2, Ahluwalia3}.

The spinor and its dual satisfy the following orthonormality relations \cite{Ahluwalia2}:
\begin{eqnarray}
\stackrel{\neg}{\lambda}_{\beta'}^{S}({\bf k})  \lambda_{\beta}^{S}({\bf k}) &=& 2 m \delta_{\beta \beta'}\,, \nonumber\\
\stackrel{\neg}{\lambda}_{\beta'}^{A}({\bf k})  \lambda_{\beta}^{A}({\bf k}) &=& -2 m \delta_{\beta \beta'}\,,\nonumber\\
\stackrel{\neg}{\lambda}^{S}_{\beta'}({\bf k})\lambda_{\beta}^{A}({\bf k})&=&\stackrel{\neg}{\lambda}^{A}_{\beta'}({\bf k})\lambda_{\beta}^{S}({\bf k})=0.\label{preserve}
\end{eqnarray}

In contrast to the bosonic scalar field, the creation and annihilation operators for the Elko field satisfy anti-commutation relations like fermions \cite{Ahluwalia2}, namely\begin{eqnarray}
\left\{ a_{\beta,n}({\bf k}),a^{\dagger}_{\beta',n'}({\bf k'})\right\}&=&(2\pi)^{2}\delta^{(2)}({\bf k}-{\bf k}')\delta_{\beta\beta'}\delta_{nn'}\\
\left\{ a_{\beta,n}({\bf k}),a_{\beta',n'}({\bf k'})\right\}&=&\left\{ a^{\dagger}_{\beta,n}({\bf k}),a^{\dagger}_{\beta',n'}({\bf k'})\right\}=0.
\end{eqnarray}

The Hamiltonian operator $\hat{H}$ for the Elko field is
\ie
\hat{H}=\sum_{n=1}^{\infty}\int \frac{d^{2}k}{(2\pi)^{2}}2\omega_{n,\xi}({\bf k})\sum_{\beta}\left(a^{\dagger}_{\beta,n}({\bf k})a_{\beta,n}({\bf k})-a_{\beta,n}({\bf k})a^{\dagger}_{\beta,n}({\bf k})\right), \label{Hamiltonian-E}
\fe and the corresponding vacuum energy for the Elko field may be written as 
\ie
E^{(Elko)}=-4 \  E^{(scalar)}=-4 \ \frac{\ell^{\xi-1}L^{2}}{(2\pi)^{2}}\int d^{2}k\sum_{n=1}^{\infty}\frac{1}{2}\sqrt{\mu^{2\xi}+\left(k_{x}^{2}+k_{y}^{2}+\left(\frac{n\pi}{a}\right)^{2}\right)^{\xi}}.
\label{vaccum2e}\fe

As can be seen from equation \eqref{vaccum2e} and  \eqref{vaccum2}, the difference between the vacuum energies of the Elko field and the real scalar field lies on the -4 factor in front of the zero-point energy. The minus sign reminds us of the fermionic character of the Elko field, and the factor of 4 refers to the fact that all the four degrees of freedom associated with $\eta(x)$ carry a vacuum energy $=-(1/2)\omega({\bf k})$ \cite{Ahluwalia3,Ahluwalia2}.

\section{The vacuum energy in a Lifshitz-like field theory}\label{sec-a}

The vacuum energy represented by the integral \eqref{vaccum2} is infinite, and so some regularization schemes must be employed to remove this divergence. Several regularization methods can be implemented to calculate the Casimir energy; for example, the adoption of a well-behaved cutoff was used in Ref. \cite{Petrov} to determine the Casimir energy for the massless scalar field in the {LP} scenario. In  Ref. \cite{Petrov2}, the authors use the Abel-Plana formula in order to obtain the Casimir energy in this same context. In Ref. \cite{Pereira} the Casimir energy for the Elko field was determined using the Poisson sum formula. A technique widely used in the general context of quantum field theories is the so-called dimensional regularization \cite{giambiagi}, based on the analytical continuation in the spatial dimension number. This regularization scheme allows us to observe the behavior of the Casimir force concerning the dimension of the transverse space and leads to finite energies without any explicit subtraction \cite{Amb,B1, KimballBook}. In what follows, we will use the dimensional regularization to evaluate the integral \eqref{vaccum2} in different cases involving the mass $m$ and the {LP} critical exponent $\xi$.

\subsection{Casimir energy and {Lifshitz fixed point} modifications: massless case}\label{sec-massless}

As the most straightforward case for the calculation of the Casimir energy modified by the {LP} critical exponent, let us consider a real massless scalar field. In this case, the dimensionally regularized integral \eqref{vaccum2} takes the form
\ie
E_{\xi}^{(reg)}=\ell^{\xi-1}L^{d}\sum_{n=1}^{\infty}\int\frac{d^{d}k}{(2\pi)^{d}}\frac{1}{2}\left[k^{2}+\left(\frac{n\pi}{a}\right)^{2}\right]^{\xi/2},\label{RegE1} 
\fe where $d$ is the transverse dimension assumed as a continuous, complex variable. The term in brackets under the integral can be expressed conveniently through the Schwinger proper-time representation:
\ie
\frac{1}{a^{z}}=\frac{1}{\Gamma(z)}\int_{0}^{\infty}dt~t^{z-1}e^{-at}.
\fe

After performing the moment integration, the equation \eqref{RegE1} becomes
\ie
\mathcal{E}_{\xi}^{(reg)}=\frac{\ell^{\xi-1}}{2(4\pi)^{\frac{d}{2}}\Gamma\left(-\frac{\xi}{2}\right)}\sum_{n=1}^{\infty}\int_{0}^{\infty}dt~ t^{-\frac{(d+\xi+2)}{2}}e^{-\left(\frac{n\pi}{a}\right)^{2}t},
\fe where $\mathcal{E}_{\xi}^{(reg)}$ is the regularized energy density between the plates. The remaining integral can be made using the Euler Representation for {the} gamma function, and the summation in $n$ is carried out employing the definition of the Riemann zeta function. Therefore, the energy density is expressed as
\ie
\mathcal{E}_{\xi}^{(reg)}=\frac{\ell^{\xi-1}}{2(4\pi)^{\frac{d}{2}}\Gamma\left(-\frac{\xi}{2}\right)}\left(\frac{\pi}{a}\right)^{d+\xi}\Gamma\left(-\frac{(d+\xi)}{2}\right) \zeta(-d-\xi).
\label{Ed1}\fe
We note that this expression is indeterminate for $d+\xi$ positive integer even. Nevertheless, we can apply the reflection formula \cite{Amb, KimballBook, Arfken}
\ie
\Gamma\left(\frac{z}{2}\right)\zeta(z)\pi^{-\frac{z}{2}}=\Gamma\left(\frac{1-z}{2}\right)\zeta(1-z)\pi^{\frac{z-1}{2}}, \label{reflection}
\fe and rewrite the energy density \eqref{Ed1} as
\ie
\mathcal{E}_{\xi}^{(reg)}=\frac{\ell^{\xi-1}}{2^{d+1}a^{d+\xi}\pi^{\frac{d+1}{2}}\Gamma\left(-\frac{\xi}{2}\right)}\Gamma\left(\frac{d+\xi+1}{2}\right)\zeta(d+\xi+1),\label{EnergyMassless}
\fe which is finite for every $d$ and $\xi$ positive integer.

A case of particular interest is when $d = 2$, which implies a Casimir energy per unit area
\ie
\mathcal{E}^{(\mbox{cas})}_{\xi}\equiv\frac{E_{\xi}^{(\mbox{cas})}}{L^{2}}=-\frac{\ell^{\xi-1}}{2^{\xi+4}a^{\xi+2}\pi^{2}}\sin\left(\frac{\pi\xi}{2}\right)\Gamma(\xi+2)\zeta(\xi+3), \label{energy00}
\fe which corresponds exactly to the result found in Ref. \cite{Petrov2} obtained via Abel-Plana formula. In particular, we note that the Casimir energy is always zero for $\xi$ even, and it is changing the signal to $\xi$ odd.  {The origin of the cancellation for $\xi=2n$ with $n$ positive integer follows immediately from the application of the Schwinger proper-time representation, which introduces a factor of $\Gamma(-n)$ into the denominator of the regularized energy density, as can be seen in \eqref{EnergyMassless}. This result is similar to the well-known cancellation of the massless Feynman integrals associated with so-called tadpole diagrams \cite{Regularization-Rev}.} 

Once that the Casimir force per unit of area (Casimir pressure) is obtained by $\mathcal{F}_{\xi}=-\frac{\partial \mathcal{E}_{\xi}}{\partial a}$, let us explicit some values of $\xi$ for the Casimir energy and force, considering $d=2$ in Table \ref{Table-xi}. 
\begin{table}[!htb]
\centering
\begin{tabular}{|c|c|c|c|c|c|c|}
\hline 
Field & $ $ & $\xi=1$&  $\xi=2$ & $\xi=3$ & $\xi=4$ & $\xi=5$\\
\hline
\quad Scalar \quad  & $\mathcal{E}_{\xi}^{(\mbox{cas})}$ & $-\frac{\pi^{2}}{1440a^{3}}$ & $\quad 0 \quad$ & $+\frac{\pi^{4}\ell^{2}}{5040a^{5}}$& $\quad 0 \quad $&  $-\frac{\pi^{6}\ell^{4}}{6720a^{7}}$\\
&$\mathcal{F}^{(\mbox{cas})}_{\xi}$ & $-\frac{\pi^{2}}{480a^{4}}$ & $\quad 0 \quad$ & $+\frac{\pi^{4}\ell^{2}}{1008a^{6}}$& $\quad 0 \quad $&  $-\frac{\pi^{6}\ell^{4}}{960a^{8}}$\\
\hline
Elko & $\mathcal{E}_{\xi}^{(\mbox{cas})}$ & $+\frac{\pi^{2}}{360a^{3}}$ & $\quad 0 \quad$ & $-\frac{\pi^{4}\ell^{2}}{1260a^{5}}$& $\quad 0 \quad $&  $+\frac{\pi^{6}\ell^{4}}{1680a^{7}}$\\
& $\mathcal{F}^{(\mbox{cas})}_{\xi}$ & $+\frac{\pi^{2}}{120a^{4}}$ & $\quad 0 \quad$ & $-\frac{\pi^{4}\ell^{2}}{270a^{6}}$& $\quad 0 \quad $&  $+\frac{\pi^{6}\ell^{4}}{240a^{8}}$ \\
\hline
\end{tabular}
\caption{The Casimir energy per unit of area and the Casimir force per unit of area for the massless real scalar field and the massless Elko considering some values of critical exponent $\xi$ and $d=2$.}
\label{Table-xi}
\end{table}

From Table \ref{Table-xi} we note that the results are in agreement with the massless scalar field in LP field theory presented in Ref. \cite{Petrov, Petrov2}. In particular, for  $\xi=1$ we have the  well-know case for the massless real scalar field in $(3+1)$ dimensions  $\left(\mathcal{F}_{1}^{(scalar)}=-\frac{\pi^{2}}{480a^{4}}\right)$ \cite{KimballBook}. Moreover, the Elko field with $\xi=1$ and small mass agrees with the result found in Ref. \cite{Pereira}, where $\left(\mathcal{E}_{1}^{(\mbox{cas})}=+\frac{\pi^2}{360a^3}-\frac{m^2}{24a}\right)$.

\subsection{Casimir energy and  {Lifshitz fixed point} modifications: massive case}\label{sec-mass}

For the case of a massive real scalar field, the Casimir energy modified by the {LP} critical exponent $\xi$ is given by
\begin{eqnarray}
\mathcal{E}_{\xi}^{(reg)}(m)= \frac{\ell^{\xi-1}}{\Gamma\left(-\frac{1}{2}\right)\Gamma\left(\frac{d}{2}\right)(4\pi)^{d/2}}\sum_{n=1}^{\infty}\int_{0}^{\infty}dt~t^{-3/2}e^{-(\mu^{2\xi})t}\int_{0}^{\infty}dk~k^{d-1}e^{-\left(k^{2}+\left(\frac{n\pi}{a}\right)^{2}\right)^{\xi}t},\nonumber\\
\label{MassiveEnergy}
\end{eqnarray} where $\mu^{\xi}=m\ell^{1-\xi}$. In this general case, the integration over $k$ does not have a closed-form in terms of known functions. Therefore, we will restrict ourselves to evaluating the exact form of \eqref{MassiveEnergy} for some particular values of $\xi$.
 
Assuming $\xi = 1$ and proceeding in a similar way to the massless case, the equation \eqref{MassiveEnergy} takes the form
\ie
\mathcal{E}_{1}^{(reg)}(m)=-\frac{1}{4\sqrt{\pi}(4\pi)^{\frac{d}{2}}}\Gamma\left(-\frac{(d+1)}{2}\right) \sum_{n=1}^{\infty}\left(m^{2}+\frac{n^{2}\pi^{2}}{a^{2}}\right)^{\frac{d+1}{2}}.
\label{MassiveSum}
\fe 

We can perform the summation using the functional relation {(an Epstein-Hurwitz Zeta function type)} \cite{Amb,chineses}
\begin{equation}
\sum_{n=-\infty}^{\infty} \left (bn^2+\mu^2\right )^{-s}=\frac{\sqrt{\pi}}{\sqrt{b}}\frac{\Gamma\left (s-\frac 12\right )}{\Gamma(s)}\mu^{1-2s}
+\frac{\pi^s}{\sqrt{b}}\frac{2}{\Gamma(s)}\sum_{n=-\infty}^{\infty\hspace{0.2cm}\prime}\mu^{\frac 12-s}\left (\frac{n}{\sqrt{b}}\right )^{s-\frac 12}K_{\frac 12-s}\left( 2\pi \mu \frac{n}{\sqrt{b}}\right ),
\end{equation}
where $K_{\nu}(z)$ is the modified Bessel function, and the
prime means that the term $n=0$ has to be excluded in the sum. After some algebra, it can be shown that the expression \eqref{MassiveSum} results in
\begin{eqnarray}
\mathcal{E}_{1}^{(reg)}(m)	&=&	-\frac{1}{2(4\pi)^{\frac{d+2}{2}}}\left\{ -\sqrt{\pi}\Gamma\left(-\frac{d+1}{2}\right)m^{d+1}\right.\nonumber\\
		&+&\left.   a m^{d+2}\Gamma\left(-\frac{d+2}{2}\right)+\frac{4m^{\frac{d+2}{2}}}{a^{\frac{d}{2}}}\sum_{n=1}^{\infty}n^{-\frac{d+2}{2}} K_{\frac{d+2}{2}}\left(2amn\right)\right\}.
\end{eqnarray}
The first term in brackets does not depend on $a$, so it does not contribute to the Casimir force. The second term in brackets depends linearly on the distance between the plates and produces a constant Casimir force. It is, in fact, related to the Casimir energy of the vacuum in the absence of the plates (being discarded). The third term in brackets is the relevant part for the Casimir energy. Thus, we find the following Casimir energy density for $\xi = 1$:
\ie
\mathcal{E}_{1}^{(reg)}(m)=-2\left(\frac{am}{4\pi}\right)^{\frac{d+2}{2}}\frac{1}{a^{d+1}}\sum_{n=1}^{\infty}n^{-\frac{d+2}{2}}K_{\frac{d+2}{2}}\left(2amn\right),\label{Massiveenergy2}
\fe which recovers the classical result for the Casimir energy for Dirichlet condition \cite{KimballBook}. For completeness, let us evaluate the general result \eqref{Massiveenergy2} at $d = 2$. This expression implies a Casimir energy 
\ie
E_{1}^{(\mbox{cas})}(m)=-\frac{L^{2}}{8\pi^{2}}\frac{m^{2}}{a}\sum_{n=1}^{\infty}\frac{1}{n^{2}}K_{2}\left(2amn\right),\label{MassiveEnergy3}
\fe whose asymptotic behavior for $m\ll a^{-1}$ gives
\ie
E_{1}^{(\mbox{cas})}(m\ll a^{-1})= -\frac{L^{2}\pi^{2}}{1440}\frac{1}{a^{3}}+\frac{L^{2}m^{2}}{96}\frac{1}{a}+\cdots, \label{massa}
\fe so that the first term corresponds to the Casimir energy for the usual massless scalar field.  {Note that the mass term decreases the massless Casimir energy.}

On the other extreme, i.e., for $m\gg a^{-1}$, the Casimir energy \eqref{MassiveEnergy3} decays exponentially with the mass of the particle, namely
\ie
E_{1}^{(\mbox{cas})}(m\gg a^{-1})= -\frac{L^{2}}{16\pi^{2}}\frac{m^{2}}{a}\left(\frac{\pi}{ma}\right)^{1/2}e^{-2ma},
\fe leading to a small force at the non-relativistic limit \cite{KimballBook, Report1986}.

Now, we consider the massive case when the {LP} critical exponent assumes the value of $\xi = 2$. In this case, the integrations over $k$ and on the $t$ parameter can still be performed analytically. We can show that, for this case, the Eq. \eqref{MassiveEnergy} results in
\begin{eqnarray}
\mathcal{E}_{2}^{(reg)}(m)&=& \frac{\ell\sec\left(\frac{\pi d}{2}\right)}{2^{\frac{3 d}{2}+2} \pi^{\frac{3 d-1}{2}}}\sum_{n=1}^{\infty} \left(\frac{a}{n}\right)^d \left(\mu^{4}+\frac{\pi^{4}n^{4}}{a^{4}}\right)^{\frac{d+1}{2}}\!\! \,_2\tilde{F}_1\left(\frac{d}{4},\frac{d+2}{4};\frac{d+3}{2};1+\frac{a^4 \mu ^4}{n^4 \pi ^4}\right)\nonumber\\
&-&\frac{d \ell \Gamma\left(\frac{d+5}{2}\right)\Gamma(-d-3)}{(2\pi)^{-\frac{d}{2}-2}\Gamma\left(\frac{d}{2}+1\right)}\sum_{n=1}^{\infty}\left(\frac{n}{a}\right)^{d+2}\!\! \,_2\tilde{F}_1\left(\frac{-d-2}{4},-\frac{d}{4};\frac{1-d}{2};1+\frac{a^4 \mu ^4}{n^4 \pi ^4}\right),\nonumber\\\label{Energyzeta2}
\end{eqnarray}
where $\mu^{4}=m^{2}\ell^{-2}$, and $\,_{2}\tilde{F}_{1}(a,b,c;z)$ is the regularized hypergeometric function, defined by $\,_{2}F_{1}(a,b,c;z)/\Gamma(c)$.

Unfortunately, there is no closed expression for the sum over $n$ in \eqref{Energyzeta2}. So, in order to get some information about the behavior of the $\mathcal{E}_{2}^{(reg)}(m)$, let us  expand \eqref{Energyzeta2} around $a = 0$. Thus, the leading order contribution to the Casimir energy is mass-independent and can be written as 
\ie
\mathcal{E}_{2}^{(reg)}=\frac{\ell\pi^{\frac{d}{2}+2}\sec\left(\frac{\pi  d}{2}\right)}{2^{d+1} a^{d+2} \Gamma \left(\frac{d+4}{2}\right)}\zeta(-d-2)-\frac{8 \ell \pi ^{\frac{d+1}{2}} \sin \left(\frac{\pi  d}{2}\right)}{ a^{d+2}} \zeta (-d-2) \Gamma (-d-3) \Gamma \left(\frac{d+5}{2}\right),
\fe and it can be shown that this result is identically null. Therefore, the Casimir energy $\mathcal{E}_{2}^{(reg)}$ is zero in this approximation, agreeing that the result previously obtained for the massless case (Table \ref{Table-xi}).

\subsection*{The small-mass regime with arbitrary $\xi$}

Now we will consider the small-mass regime which we assume $\mu^{\xi}\equiv m\ell^{1-\xi}\ll a^{-1}$, such that we can write
\ie
\left[\mu^{2\xi}+\left(k^{2}+\left(\frac{n\pi}{a}\right)^{2}\right)^{\xi}\right]^{\frac{1}{2}}=\left[k^{2}+\left(\frac{n\pi}{a}\right)^{2}\right]^{\frac{\xi}{2}}+\frac{\mu^{2\xi}}{2}\left[k^{2}+\left(\frac{n\pi}{a}\right)^{2}\right]^{-\frac{\xi}{2}}+\mathcal{O}(\mu^{2\xi}).
\label{Massaprox}\fe
Substituting the above expression in \eqref{vaccum2}, the first term in \eqref{Massaprox} results in the Casimir energy for the massless scalar field identical to that obtained in \eqref{EnergyMassless}. The second term gives rise to correction due to mass and can be put into the form
\ie
\Delta\mathcal{E}_{\xi}^{(reg)}(m)=\frac{m^{2}\ell^{1-\xi}}{2^{d+2}\pi^{\frac{d+1}{2}}a^{d-\xi}\Gamma\left(\frac{\xi}{2}\right)}\Gamma\left(\frac{d-\xi+1}{2}\right)\zeta\left(d-\xi+1\right).\label{MassAprox2}
\fe

For the case of interest, $d = 2$, the mass correction to the Casimir energy becomes
\begin{eqnarray}
\Delta\mathcal{E}_{\xi}(d=2,m)&=&\frac{m^{2}\ell^{1-\xi}}{2^{5-\xi}\pi^{2}a^{2-\xi}}\sin\left(\frac{\pi\xi}{2}\right)\Gamma\left(2-\xi\right)\zeta\left(3-\xi\right)\nonumber \\
&=&\frac{m^2 \pi ^{1-\xi } a^{\xi -2} \ell ^{1-\xi }}{8 (\xi -2)}\zeta (\xi -2), \label{eee2}
\end{eqnarray}
where we use the reflection \eqref{reflection} and the Legendre duplication formula $\Gamma(s)=2^{s-1}\pi^{-1/2}\Gamma(\frac{s}{2})\Gamma(\frac{s+1}{2})$. Considering $\xi>0$, the equation \eqref{eee2} diverge only for $\xi=2$ and $\xi=3$. To isolate these divergences and determine the relevant part of Casimir energy, let us assume $d = 2-\epsilon$ in \eqref{MassAprox2} and expand around $\epsilon = 0$.

For $\xi=2$, one finds
\ie
\Delta\mathcal{E}_{2}^{(reg)}(d=2-\epsilon,m)=-\frac{m^2}{16 \pi \ell}\frac{1}{\epsilon}+\frac{m^2 }{32\pi\ell}\left(\gamma -\ln \left(16 \pi  a^2\right)\right)+\mathcal{O}(\epsilon),
\fe where $\gamma$ is the Euler-Mascheroni constant, with numerical value $\gamma\approx 0.577216$. We observe that the term containing the pole at $\epsilon = 0$ is independent of $a$, so it does not contribute to the Casimir force. The finite contribution and $a$ dependent, at the limit $\epsilon\rightarrow 0$, is then
\ie
\Delta\mathcal{E}_{2}^{(\mbox{cas})}(m)=-\frac{m^2}{16 \pi \ell}\ln a.\label{DeltaMass1}
\fe

For $\xi=3$, one finds
\ie
\Delta\mathcal{E}_{3}^{(reg)}(d=2-\epsilon,m)=\frac{a m^{2}}{8\pi^{2}\ell^{2}}\frac{1}{\epsilon}+\frac{a m^{2}}{16\pi^{2}\ell^{2}}\left(\gamma+\ln\left(\frac{a^2}{\pi}\right)\right)+\mathcal{O}(\epsilon),
\fe where we note that the divergent part is linear with respect to the parameter $a$, and it is related to the vacuum energy in the absence of the plates (being discarded). The relevant part is then
\ie
\Delta\mathcal{E}_{3}^{(\mbox{cas})}(m)= \frac{a m^2 }{8 \pi ^2 \ell^2}\ln a.\label{DeltaMass2}
\fe

Table \ref{Table-2} shows the small-mass corrections for the densities of the Casimir energy and the Casimir force of real massive scalar field for some values of critical exponent $\xi$ with $d=2$. The Casimir effect for the Elko field is embedded. We note that, in the small-masses regimen, the Casimir forces for $\xi\geqslant 3$ increase with the distance between plates $a$,  {which is an unusual behavior since in the ordinary case ($\xi = 1$) it vanishes when the distance $a$ tends to grow. This result becomes relevant, especially when $\xi$ is even, such that the correction due to the mass term becomes dominant, as seen in the previous massless case. Figure \ref{Fig-m} shows the density of the Casimir force for $\xi=1$ with some masses parameters (derived from the equations \eqref{energy00} and \eqref{MassiveEnergy3}). Note the decreasing of Casimir force with the mass parameter. Moreover, Figure \ref{Fig-m} also shows the variation of force with the $\xi$ parameter for the small-masses regimen with $m=10^{-2}$.

{The above result for the Casimir effect was obtained considering only free theories. However, there are other relevant deformations with several numbers of spatial derivatives that can be generated at the quantum level when the interactions are turned on. It would be interesting to understand the effects of these relevant deformations on the Casimir effects since they usually dominate over the mass term around a Lifshitz fixed point with $\xi >1$. Establishing the physical implications of the radiative corrections induced by these terms is an interesting open issue for future investigation.}

\begin{table}[!htb]
\centering
\begin{tabular}{|c|c|c|c|c|c|c|}
\hline 
Field& $ $ &$\xi=1$&  $\xi=2$ & $\xi=3$ & $\xi=4$ & $\xi=5$\\ 
\hline
Scalar& $\Delta\mathcal{E}_{\xi}^{(\mbox{cas})}$ & $\frac{m^{2}}{96a}$ & $\quad -\frac{m^2}{16 \pi \ell^2}\ln a \quad$ & $\quad \frac{a m^2 }{8 \pi ^2 \ell^2}\ln a \quad$& $\frac{a^2 m^2}{96 \pi  \ell ^3} $&  $\quad \frac{a^3 m^2 \zeta (3)}{24 \pi ^4 \ell ^4} \quad$ \\ 
& $\Delta\mathcal{F}_{\xi}^{(\mbox{cas})}$ & $\frac{m^2}{96 a^2}$ & $\frac{m^2}{16 \pi \ell^2  a }$ & $\frac{m^2 (1+\ln a)}{8 \pi  \ell^2}$& $-\frac{a m^2}{48 \pi  \ell^3}$&  $-\frac{a^2 m^2 \zeta (3)}{8 \pi ^4 \ell^4}$\\
\hline
Elko& $\Delta\mathcal{E}_{\xi}^{(\mbox{cas})}$ & $-\frac{m^{2}}{24a}$ & $\quad \frac{m^2}{4 \pi \ell^2}\ln a \quad$ & $\quad- \frac{a m^2 }{2 \pi ^2 \ell^2}\ln a \quad$& $-\frac{a^2 m^2}{24 \pi  \ell ^3} $&  $\quad -\frac{a^3 m^2 \zeta (3)}{6 \pi ^4 \ell ^4} \quad$\\ 
& $\Delta\mathcal{F}_{\xi}^{(\mbox{cas})}$ & $-\frac{m^2}{24 a^2}$ & $-\frac{m^2}{4 \pi \ell^2  a }$ & $-\frac{m^2 (1+\ln a)}{2 \pi  \ell^2}$& $\frac{a m^2}{12 \pi  \ell^3}$&  $\frac{a^2 m^2 \zeta (3)}{2 \pi ^4 \ell^4}$\\
\hline
\end{tabular}
\caption{The small-mass corrections for the Casimir energy (force) per unit of area for the massive scalar field and the massive Elko field with some values of critical exponent $\xi$ and $d=2$.}
\label{Table-2}
\end{table}


\begin{figure}[!hbt] 
\begin{minipage}[t]{0.48 \linewidth}
   \includegraphics[width=\linewidth]{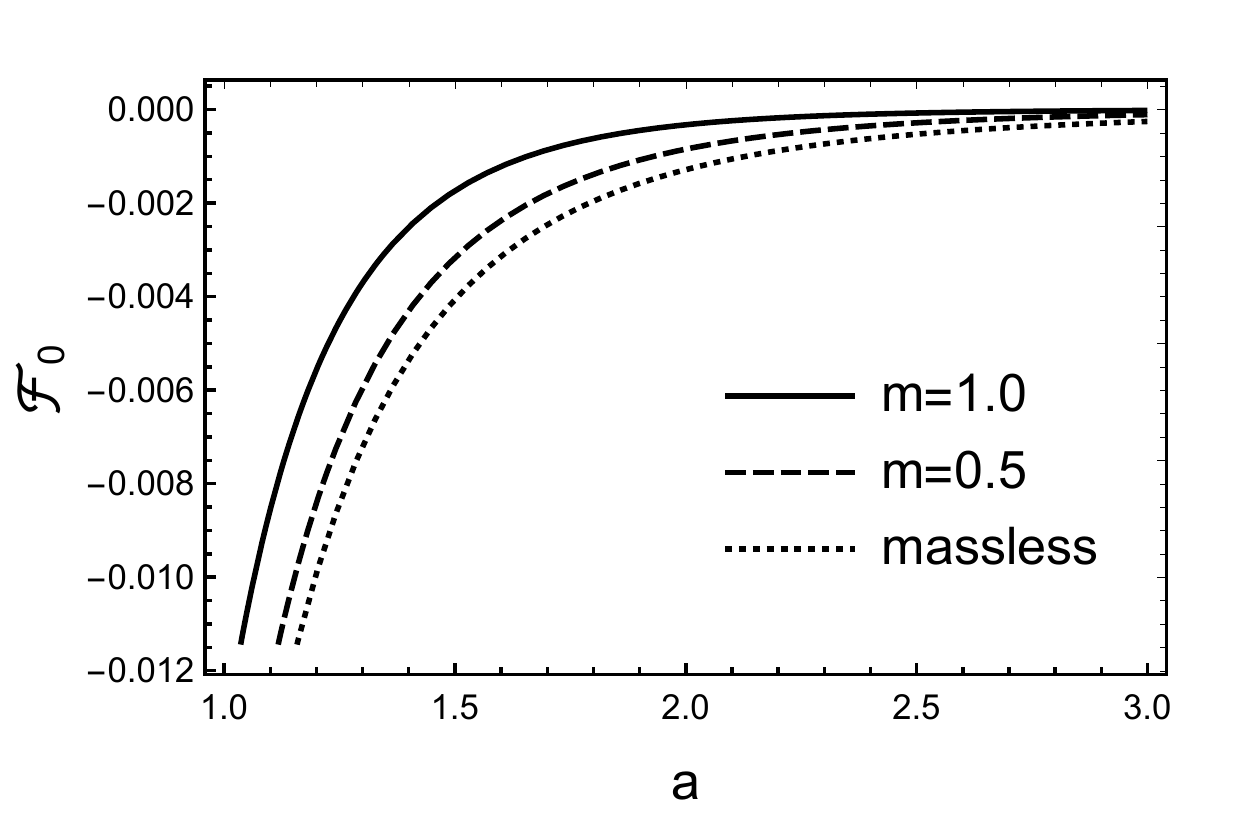}                             
\end{minipage}
\,
\begin{minipage}[t]{0.48 \linewidth}
    \includegraphics[width=\linewidth]{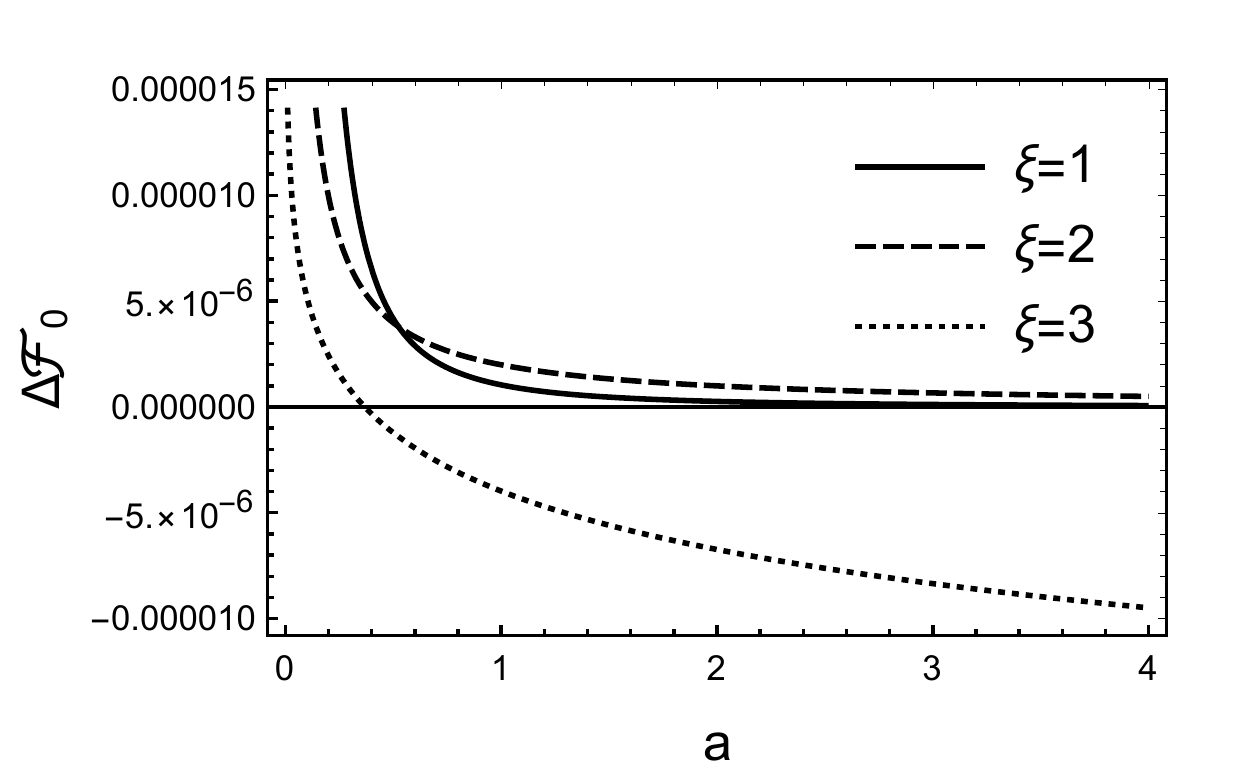}

\end{minipage}
\caption{Casimir force for the real scalar field for $d=2$ and $\ell=1$. Dependence of mass on the force with $\xi=1$ (left figure). Dependence of $\xi$ parameter on the correction of the force with $m=10^{-2}$ (right figure).}
 \label{Fig-m}
\end{figure}

\section{Thermal corrections and the role of the  {Lifshitz fixed point parameter}}\label{sec-temp}

In this section, we study the thermal corrections to the Casimir energy for the massless case of the scalar field in a  field theory at a Lifshitz fixed point. Where we bring attention to the case of interest $d = 2$ and $\xi$ arbitrary. In this context, a great read for the general applications of finite-temperature and some useful formulas  can be found in Ref. \cite{Kapusta}. 

Following the approach described in Refs.  \cite{KimballBook,Bordag2, Report1986} for the Casimir effect, the Casimir free energy $F_{C}$ can be decomposed in the form
\begin{equation}
F_{C}=E^{(0)}_{C}+\tilde{F}_{C},\label{FreeEnergy}
\end{equation} where the $E^{(0)}_{C}$ is the zero temperature energy contribution (previously computed in Eq. (\ref{EnergyMassless}) and $\tilde{F}_{C}$ is the temperature-dependent part of the Casimir free energy which has the form \cite{KimballBook, Bordag2}:
\begin{equation}
\tilde{F}_{C}(a,T,\xi)=\frac{\kappa_{B}TL^{2}}{2}\int\frac{d^{2}k}{(2\pi)^{2}}\sum_{n=-\infty}^{+\infty}\ln\left(1-e^{-\beta\omega_{n,\xi}({\bf k})}\right),\label{FreeEnergy1}
\end{equation} where $\omega_{n,\xi}({\bf k})=\ell^{\xi-1}\left[k^{2}+\left(\frac{n\pi}{a}\right)^{2}\right]^{\xi/2}$,
$\beta=(\kappa_{B}T)^{-1}$, being $T$ the temperature and $\kappa_{B}$
the Boltzmann constant. After performing the angular integration, the equation
(\ref{FreeEnergy1}) takes the form:
\begin{equation}
\tilde{F}_{C}(a,T,\xi)=\frac{\kappa_{B}TL^{2}\pi}{8a^{2}}\sum_{n=-\infty}^{+\infty}\int_{n^{2}}^{\infty}dy\ln\left(1-e^{-\tilde{\beta}y^{\frac{\xi}{2}}}\right),\label{FreeEnergy2}
\end{equation}
where we {are} changing from $k$ to the dimensionless integration variable $y=n^{2}+\left(\frac{ak}{\pi}\right)^{2}$
and introduced the constant $\tilde{\beta}=\beta\ell^{\xi-1}\left(\frac{\pi}{a}\right)^{\xi}$.
Defining the function
\begin{equation}
b(a,T,\xi,n)=\frac{\kappa_{B}T}{2}\int_{n^{2}}^{\infty}dy\ln\left(1-e^{-\tilde{\beta}y^{\frac{\xi}{2}}}\right),\label{Bfunction1}
\end{equation}
the expression (\ref{FreeEnergy2}) can be written as
\begin{equation}
\tilde{F}_{C}(a,T,\xi)=\frac{L^{2}\pi}{2a^{2}}\left\{ \frac{1}{2}b(a,T,\xi,0)+\sum_{n=1}^{+\infty}b(a,T,\xi,n)\right\}.\label{FreeEnergy3}
\end{equation}

The resulting sum in (\ref{FreeEnergy2}) cannot be performed analytically, and an exact evaluation for the thermal correction of Casimir at an arbitrary temperature is not possible. Approximate expressions for $\tilde{F}_{C}$ can be obtained in the low-temperature and in the high-temperature limit \cite{KimballBook,Bordag2, Report1986, Kapusta}.

We first consider the low-temperature regime. When $T\ll1$ $(\beta\gg1)$, a series
expansion of the logarithm in eq. (\ref{Bfunction1}) is justified.  The zero-temperature limit comes entirely from $n = 0$ :
\begin{eqnarray}
b(a,T,\xi,0) & = & \frac{\kappa_{B}T}{2}\int_{0}^{\infty}dy\left[-\sum_{m=1}^{\infty}\frac{e^{-m\tilde{\beta}y^{\frac{\xi}{2}}}}{m}\right]\nonumber \\
 & = & -\frac{\kappa_{B}T}{\xi\tilde{\beta}^{\frac{2}{\xi}}}\Gamma\left(\frac{2}{\xi}\right)\zeta\left(\frac{2+\xi}{\xi}\right).
\end{eqnarray}

The leading corrections come from $n\neq 0$, and they are characterized by exponentially small terms. Thus, we have
\begin{equation}
b(a,T,\xi,n) = -\frac{\kappa_{B}T}{\xi}\frac{e^{-\tilde{\beta}n^{\xi}}}{\tilde{\beta}n^{\xi-2}}\left[1-\left(\frac{\xi-2}{\xi}\right)\frac{1}{\tilde{\beta}n^{\xi}}+\mathcal{O}\left(\tilde{\beta}^{-2}\right)\right].
\end{equation}
Inserting these expressions into equation (\ref{FreeEnergy3}) lead to the finite-temperature correction in the low-temperature regime:
\begin{equation}
\tilde{F}_{C}(a,T,\xi)=-\frac{L^{2}\pi}{4a^{2}}\left(\frac{\kappa_{B}T}{\xi}\right)\left\{ \frac{1}{\mathbf{\tilde{\beta}^{\frac{2}{\xi}}}}\Gamma\left(\frac{2}{\xi}\right)\zeta\left(\frac{2+\xi}{\xi}\right)+2\sum_{n=1}^{+\infty}\frac{e^{-\tilde{\beta}n^{\xi}}}{\tilde{\beta}n^{\xi-2}}\left[1-\left(\frac{\xi-2}{\xi}\right)\frac{1}{\tilde{\beta}n^{\xi}}\right]\right\} .
\end{equation} The dominant contribution is obtained by taking $n=1$, so we find the result
\begin{eqnarray}
\tilde{F}_{C}(a,T,\xi) & = & -\frac{L^{2}\kappa_{B}T}{4\pi\xi}\left\{ \frac{\left(\kappa_{B}T\right)^{\frac{2}{\xi}}}{\ell^{2-\frac{2}{\xi}}}\Gamma\left(\frac{2}{\xi}\right)\zeta\left(\frac{2+\xi}{\xi}\right)\right.\nonumber\\
 & + & \left.2\left(\frac{a}{\pi}\right)^{\xi-2}\frac{\kappa_{B}T}{\ell^{\xi-1}}e^{-\frac{\ell^{\xi-1}}{\kappa_{B}T}\left(\frac{\pi}{a}\right)^{\xi}}\left[1-\left(\frac{\xi-2}{\xi}\right)\left(\frac{a}{\pi}\right)^{\xi}\frac{\kappa_{B}T}{\ell^{\xi-1}}\right]\right\}.\label{FreeEnergy4}
\end{eqnarray}
We note that the thermal correction is proportional to $1/\xi$ and decreases when $\xi$ increases. For the particular case with $\xi = 1$, one recognizes that the expression (\ref{FreeEnergy4}) recovers in the usual case presented in Ref. \cite{Report1986} (less than a $1/2$ factor due to degrees of freedom).

The calculation of the high-temperature limit, i.e., $T\gg1$ ($\beta\ll1$), becomes more difficult than the low-temperature expansion because the logarithm in (\ref{Bfunction1}) can not be expanded anymore. One convenient way to carry out the series in (\ref{FreeEnergy3}) is to use the Poisson sum formula, which states that if $c(\alpha)$ is the Fourier transform of $f(x)$,
\begin{equation}
c(\alpha)=\frac{1}{2\pi}\int_{-\infty}^{+\infty}dx\ e^{i\alpha x}f(x),
\end{equation}then the following identity is verified \cite{KimballBook,Report1986}:
\begin{equation}
\sum_{n=-\infty}^{+\infty}f(n)=2\pi\sum_{n=-\infty}^{+\infty}c(2\pi n).
\end{equation}

In our specific case we can write the expression for Casimir energy as 
\begin{equation}
\tilde{F}_{C}(a,T,\xi)=\left(\frac{L\pi}{a}\right)^{2}\left\{ \frac{1}{2}c(a,T,\xi,0)+\sum_{n=1}^{\infty}c(a,T,\xi,2\pi n)\right\},\label{FreeEnergy5}
\end{equation}
where
\begin{equation}
c(a,T,\xi,\alpha)=\frac{1}{\pi}\int_{0}^{\infty}dn\cos\left(\alpha n\right)b(a,T,\xi,n).
\end{equation}

At this point, we would like to call attention to the fact that the
term $c(a,T,\xi,0)=1/\pi\int_{0}^{\infty}dn b(a,T,\xi,n)$ {in} (\ref{FreeEnergy5}) is connected to the boundaryless free energy . In the usual case ($\xi=1$) this contribution can be calculated directly, and gives rise to the Stefan-Boltzmann law, in which $\tilde{F}\sim V T^{4}$ and $V=L^{2}a$ {represent} the spatial volume. As our main objective is to determine the Casimir energy in the presence of constraints (e.g., external plates),  we only need to calculate the part involving the sum,  and the finite-temperature correction for free energy $\tilde{F}_{C}$ becomes
\begin{equation}
\tilde{F}_{C}(a,T,\xi)=\left(\frac{L\pi}{a}\right)^{2}\sum_{n=1}^{\infty}c(a,T,\xi,2\pi n).
\end{equation}
The function $c(a,T,\xi,\alpha)$ assumes the form
\begin{eqnarray}
c(a,T,\xi,\alpha) & = & -\frac{\kappa_{B}T}{\pi\alpha}\frac{\partial}{\partial\alpha}\left[\int_{0}^{\infty}dn\cos\left(\alpha n\right)\ln\left(1-e^{-\tilde{\beta}n^{\xi}}\right)\right]\nonumber\\
 & = & \frac{\ell^{\xi-1}\xi}{\pi\alpha}\left(\frac{\pi}{a}\right)^{\xi}\frac{\partial}{\partial\alpha}\left[\frac{1}{\alpha}\int_{0}^{\infty}dn\sin\left(\alpha n\right)\frac{n^{\xi-1}}{e^{\tilde{\beta}n^{\xi}}-1}\right],\label{Cfunction1}
\end{eqnarray} where in the first line of (\ref{Cfunction1}) we integrate by parts. To make a high-temperature expansion of $c(a,T,\xi,\alpha)$ we can use the identity 
(see appendix in Ref. \cite{Kapusta})
\begin{equation}
\frac{1}{e^{z}-1}=\frac{1}{z}-\frac{1}{2}+2\sum_{l=1}^{\infty}\frac{z}{z^{2}+(2\pi l)^{2}}.
\end{equation}
Then, it follows that
\begin{eqnarray}
c(a,T,\xi,\alpha) & = & -\frac{\kappa_{B}T\xi}{2\alpha^{3}}+\frac{\left(\ell\pi\right)^{\xi-1}}{2a^{\xi}\alpha^{\xi+3}}\Gamma\left(2+\xi\right)\sin\left(\frac{\pi\xi}{2}\right)\nonumber\\
 & + & \frac{2\kappa_{B}T\xi}{\pi\alpha}\frac{\partial}{\partial\alpha}\left[\frac{1}{\alpha}\sum_{l=1}^{\infty}\int_{0}^{\infty}dn\sin\left(\alpha n\right)\frac{n^{2\xi-1}}{n^{2\xi}+\left(\frac{2\pi l}{\tilde{\beta}}\right)^{2}}\right].\label{Cfunction2}
\end{eqnarray}

The last integral over $n$ does not have a closed analytic expression for an arbitrary $\xi$. Keeping only the first two terms in (\ref{Cfunction2}), the free energy $\tilde{F}_{C}$ in the high-temperature expansion becomes 
\begin{equation}
\tilde{F}_{C}(a,T,\xi)=-\frac{L^{2}\kappa_{B}T\xi}{16\pi a^{2}}\zeta\left(3\right)+\frac{L^{2}\ell^{\xi-1}}{2^{\xi+4}a^{\xi+2}\pi^{2}}\sin\left(\frac{\pi\xi}{2}\right)\Gamma\left(\xi+2\right)\zeta\left(\xi+3\right).\label{FreeEnergy6}
\end{equation}
We observe that the first term in the above expression recovers the leading correction at the high-temperature limit when $\xi=1$ \cite{KimballBook}. Besides, it should be noted that the second term of eq. (\ref{FreeEnergy6}) is precisely the Casimir energy at zero-temperature with opposite sign.  Therefore, it is canceled in the total Casimir free energy defined in (\ref{FreeEnergy}). This result is a general characteristic of Casimir thermal corrections at the high-temperature regime, as discussed by Ref. \cite{Report1986}.

\section{Conclusions}\label{sec-conc}

In this work, we perform the calculation of Casimir energy and force for two fields: the real scalar field and the Elko field in a field theory at a Lifshitz fixed point. We study the equation of motion for both fields and remark its differences.  Using the dimensional regularization, we obtain the expression for the Casimir effect in term of dimensional parameters $d$ and the {LP} parameter $\xi$. Our results generalize those obtained in Refs.  \cite{Petrov, Petrov2} for the massless scalar field and also for the Elko field obtained in Ref. \cite{Pereira}. For the massive case, we obtain the usual expressions for $\xi=1$ where the mass decrease{s} slightly the Casimir force. With $\xi=2$, we observe that the Casimir energy still null, regardless of mass value. In the small-mass regime the Casimir energy depends on $m^2$ for any $\xi$,  however, the Casimir force is unusually increasing with the distance for $\xi \geq 3$. Besides, we study the thermal correction to the Casimir effect for the low-temperature and the high-temperature limits. At both limits, the {LP} parameter modifies the usual results. As perspectives, we want to study the scalar field with a similar {LP} anisotropy in the context of cosmology, cosmic inflation, and braneworlds models.


\section*{Acknowledgments}
\hspace{0.5cm}The authors would like to thank the Funda\c{c}\~{a}o Cearense de apoio ao Desenvolvimento Cient\'{\i}fico e Tecnol\'{o}gico (FUNCAP), the Coordena\c{c}\~ao de Aperfei\c{c}oamento de Pessoal de N\'ivel Superior (CAPES), and the Conselho Nacional de Desenvolvimento Cient\'{\i}fico e Tecnol\'{o}gico (CNPq) for financial support. R. V. Maluf and C. A. S. Almeida thank CNPq grants 307556/2018-2 and 308638/2015-8
for supporting this project.



\begin{thebibliography}{99}


\bibitem{Casimir} H. B. G. Casimir, Proc. K. Ned. Akad. Wet., {\bf 51}
793, (1948).


\bibitem{Bordag} M.~Bordag, U.~Mohideen and V.~M.~Mostepanenko, Phys. Rept. {\bf 353}, 1 (2001).
 
\bibitem{Bordag2} M.~Bordag, G.~L.~Klimchitskaya, U.~Mohideen and V.~M.~Mostepanenko, Int.\ Ser.\ Monogr.\ Phys.\  {\bf 145}, 1 (2009).
  
  
\bibitem{Casimir-exp0} M.~J.~Sparnaay, Physica {\bf 24}, 751 (1958).  
  
  
\bibitem{Casimir-exp} S.~K.~Lamoreaux, Phys.\ Rev.\ Lett.\  {\bf 78}, 5 (1997) Erratum: [Phys.\ Rev.\ Lett.\  {\bf 81}, 5475 (1998)].
  
  
\bibitem{Casimir-exp2} G.~L.~Klimchitskaya, U.~Mohideen and V.~M.~Mostepanenko, Rev.\ Mod.\ Phys.\  {\bf 81}, 1827 (2009).
  

\bibitem{Casimir-exp3} K.~Milton and I.~Brevik, Symmetry {\bf 11}, 201 (2019).
  

\bibitem{c1} C.~R.~Muniz, M.~O.~Tahim, M.~S.~Cunha and H.~S.~Vieira, JCAP {\bf 1801}, 006 (2018).
  
\bibitem{c2} M.~S.~Cunha, C.~R.~Muniz, H.~R.~Christiansen, V.~B.~Bezerra, Eur.\ Phys.\ J.\ C {\bf 76}, 512 (2016). 
  
\bibitem{Stokes} A.~Stokes and R.~Bennett, Annals Phys.\  {\bf 360}, 246 (2015).
  
\bibitem{Scalar} S.~Mobassem, Mod.\ Phys.\ Lett.\ A {\bf 29}, 1450160 (2014).
  


\bibitem{Pereira} S.~H.~Pereira, J.~M.~Hoff da Silva and R.~dos Santos, Mod.\ Phys.\ Lett.\ A {\bf 32}, 1730016 (2017).

\bibitem{Ponton} E.~Ponton and E.~Poppitz, JHEP {\bf 0106}, 019 (2001).


\bibitem{MD} 
  L.~Buoninfante, G.~Lambiase, L.~Petruzziello and A.~Stabile,
  Eur.\ Phys.\ J.\ C {\bf 79}, no. 1, 41 (2019)


\bibitem{LV} 
  M.~Blasone, G.~Lambiase, L.~Petruzziello and A.~Stabile,
  Eur.\ Phys.\ J.\ C {\bf 78}, no. 11, 976 (2018)
 
\bibitem{Petrov} A.~F.~Ferrari, H.~O.~Girotti, M.~Gomes, A.~Y.~Petrov and A.~J.~da Silva, Mod.\ Phys.\ Lett.\ A {\bf 28}, 1350052 (2013).
  
\bibitem{Petrov2} I.~J.~Morales Ulion, E.~R.~Bezerra de Mello and A.~Y.~Petrov, Int.\ J.\ Mod.\ Phys.\ A {\bf 30}, 1550220 (2015).
  
\bibitem{c3} C.~R.~Muniz, V.~B.~Bezerra and M.~S.~Cunha, Phys.\ Rev.\ D {\bf 88}, 104035 (2013).



\bibitem{Petrov3}  M.~B.~Cruz, E.~R.~Bezerra De Mello, A.~Y.~Petrov,  Mod.\ Phys.\ Lett.\ A {\bf 33}, 1850115 (2018).
 


\bibitem{Amb} J. Ambjorn and S. Wolfram, Ann. Phys. {\bf 147} (1983).

\bibitem{B1} K.~A.~Milton, ``Dimensional and dynamical aspects of the Casimir effect: Understanding the reality and significance of vacuum energy,'' hep-th/0009173.  

\bibitem{KimballBook} K.~A.~Milton, The Casimir effect: Physical manifestations of zero-point energy,  (River Edge, USA, World Scientific, 2001).

\bibitem{chineses} X.~h.~Zhai, X.~z.~Li and C.~J.~Feng, Mod.\ Phys.\ Lett.\ A {\bf 26}, 669 (2011).

\bibitem{Report1986} G.~Plunien, B.~Muller and W.~Greiner, Phys.\ Rept.\  {\bf 134}, 87 (1986). 
 
 


\bibitem{Ahluwalia1} D.~V.~Ahluwalia and D.~Grumiller, 
  Phys.\ Rev.\ D {\bf 72}, 067701 (2005).


\bibitem{Ahluwalia2} 
  D.~V.~Ahluwalia,
  Adv.\ Appl.\ Clifford Algebras {\bf 27}, 2247 (2017).

\bibitem{Ahluwalia3} D. V. Ahluwalia and D. Grumiller, JCAP {\bf 07}, 012 (2005).



\bibitem{elko-rot} 
  D.~V. Ahluwalia and S.~Sarmah,
  EPL {\bf 125}, 30005 (2019).

\bibitem{2019a} 
  M.~R.~A.~Arcod\'{i}a, M.~Bellini and R.~da Rocha,
  Eur.\ Phys.\ J.\ C {\bf 79}, 260 (2019).  
  
  
  
\bibitem{elko-dk1}
 B.~Agarwal, P.~Jain, S.~Mitra, A.~C.~Nayak and R.~K.~Verma,
  Phys.\ Rev.\ D {\bf 92}, 075027 (2015).
  

\bibitem{elko-dk2} 
S.~H.~Pereira and R.~S.~Costa,
  Mod.\ Phys.\ Lett.\ A {\bf 34}, no. 16, 1950126 (2019)

\bibitem{cosmo-1} 
  S.~H.~Pereira and T.~M.~Guimar\~{a}es,
  JCAP {\bf 1709}, 038 (2017).
  
\bibitem{cosmo-2} 
  S.~H.~Pereira, R.~F.~L.~Holanda and A.~P.~S.~Souza,
  EPL {\bf 120}, 31001 (2017).


\bibitem{cosmo-3}
  L.~Fabbri,
  Phys.\ Lett.\ B {\bf 704}, 255 (2011).  
  
\bibitem{vsr}
 D.~V.~Ahluwalia and S.~P.~Horvath,
  JHEP {\bf 1011}, 078 (2010).


\bibitem{hawking}
R.~da Rocha and J.~M.~Hoff da Silva,
  EPL {\bf 107}, 50001 (2014).
  
  
\bibitem{brane-1} 
  D.~M.~Dantas, R.~da Rocha and C.~A.~S.~Almeida,
  EPL {\bf 117}, 51001 (2017).

\bibitem{brane-2} 
  X.~N.~Zhou, Y.~Z.~Du, Z.~H.~Zhao and Y.~X.~Liu,
  Eur.\ Phys.\ J.\ C {\bf 78}, 493 (2018).

 
  
\bibitem{elko-lhc1} 
  A.~Alves, M.~Dias, F.~de Campos, L.~Duarte and J.~M.~Hoff da Silva,
  EPL {\bf 121}, 31001 (2018),

\bibitem{elko-lhc2} 
  M.~Dias, F.~de Campos and J.~M.~Hoff da Silva,
  Phys.\ Lett.\ B {\bf 706}, 352 (2012).




 {\bibitem{Q1a} 
  R.~Iengo and M.~Serone,
  Phys.\ Rev.\ D {\bf 81}, 125005 (2010)
\bibitem{Q1b} 
  T.~Mariz, J.~R.~Nascimento, A.~Y.~Petrov and C.~M.~Reyes,
  Phys.\ Rev.\ D {\bf 99}, no. 9, 096012 (2019)}


\bibitem{Hor} P.  Ho$\breve{{\rm r}}$ava, Phys. Rev. D \textbf{79}, 084008 (2009).

\bibitem{HC1} 
  G.~Calcagni,
  JHEP {\bf 0909}, 112 (2009).


\bibitem{HC2} 
  H.~Lu, J.~Mei and C.~N.~Pope,
  Phys.\ Rev.\ Lett.\  {\bf 103}, 091301 (2009).




\bibitem{HC3} 
  S.~Lepe and G.~Otalora,
  Eur.\ Phys.\ J.\ C {\bf 78}, 331 (2018).


\bibitem{HBH2} 
  S.~W.~Wei, J.~Yang and Y.~X.~Liu,
  Phys.\ Rev.\ D {\bf 99}, no. 10, 104016 (2019)


\bibitem{HBH} 
  J.~Xu and J.~Jing,
  Annals Phys.\  {\bf 389}, 136 (2018).



\bibitem{HGW1} 
  D.~Blas and H.~Sanctuary,
  Phys.\ Rev.\ D {\bf 84}, 064004 (2011).


\bibitem{HGW2} 
  A.~Emir G\"{u}mr\"{u}k\c{c}\"{u}o$\breve{\rm g}$lu, M.~Saravani and T.~P.~Sotiriou,
  Phys.\ Rev.\ D {\bf 97}, 024032 (2018).



\bibitem{HBB} 
  M.~A.~Anacleto, F.~A.~Brito, E.~Maciel, A.~Mohammadi, E.~Passos, W.~O.~Santos and J.~R.~L.~Santos,
  Phys.\ Lett.\ B {\bf 785}, 191 (2018).
  
\bibitem{Ryder} Lewis H. Ryder, Quantum Field Theory (Cambridge University Press; 2nd ed. 1996).

\bibitem{giambiagi} C. Bollini, J. J.  Giambiagi, Il Nuovo Cimento B {\bf 12} 20 (1972); G. 't Hooft, M. Veltman, Nuclear Physics B {\bf 44} 189 (1972).
  
\bibitem{Arfken} 
  G.~B.~Arfken,  H. J. Weber and F. E. Harris,
  Mathematical Methods for Physicists, Seventh Edition,
  (San Diego, USA:  Elsevier Science Publishing, 2012).  
 {  
\bibitem{Regularization-Rev} 
  G.~Leibbrandt,
  Rev.\ Mod.\ Phys.\  {\bf 47}, 849 (1975).  
\bibitem{Kapusta} 
J.~I.~Kapusta and C.~Gale,
Finite-temperature field theory: Principles and applications, Second Edition,
(Cambridge, UK;  University Press, 2006).}  


\end{thebibliography}
\end{document}